# Influence of surface finish and residual stresses on the ageing sensitivity of biomedical grade zirconia

Sylvain Deville, Jérôme Chevalier, and Laurent Gremillard

Associate Research Unit 5510, Materials Science Department, National Institute of Applied Science (GEMPPM-INSA), Bât B. Pascal, 20 av. A. Einstein, 69621 Villeurbanne Cedex, France

**Abstract**

We demonstrate in this paper the influence of surface finish on the ageing kinetics of biomedical grade zirconia. The critical influence of polishing has been systematically investigated by optical microscopy, atomic force microscopy and X-ray diffraction. The stress state around polishing scratches gives rise to preferential transformation in the zone of the induced scratches and consequently to accelerated ageing. The influence of residual stresses is analyzed semi-quantitatively by preparing samples with various surface finish, thus with various stress states. Rough polishing produces a compressive surface stress layer beneficial for the ageing resistance, while smooth polishing produces preferential transformation nucleation around scratches. When a thermal treatment of 2 h at 1200 °C is applied to relax the residual stresses, all the surfaces states exhibit the same sensitivity to ageing. These results demonstrate that roughness alone cannot be used for ensuring a long-term stability. The variation of ageing sensitivity is indeed related indirectly to the surface roughness via the induced surface stress state. The current ISO standards are not able to take these effects into account. Indeed, great variations in ageing kinetics were observed for samples with different surface states, although all well below the ISO requirements.

## 1. Introduction

Hard on hard materials combinations are increasingly used for the head/cup couple of orthopaedic implants. The reasoning behind this choice is to reduce wear effects and particle release usually associated with polymer (usually, ultra-high molecular weight polyethylene, UHMWPE) and metallic alloy combinations. Although alumina was the first ceramic used for the cup, its development has been, at least for the first generations of alumina implanted, limited by its inherent brittleness. Yttria-stabilized zirconia (Y-TZP), with its better mechanical properties, was introduced subsequently as an alternative. Of particular interest is the fracture toughness, which is at least twice as high as that of alumina, and a higher fatigue-crack propagation threshold, potentially ensuring a longer implant lifetime. The use of zirconia has allowed new, improved implant designs not possible with the more brittle alumina. Owing to the stabilizing effect of yttria (Y2O3), Y-TZP materials can be processed in the metastable tetragonal (t) structure. The retention of the t phase at ambient temperature allows it to transform to the monoclinic (m) structure under external applied stresses. This transformation and the resulting compressive stresses generated in the vicinity of a propagating crack, combined with the volume increase associated with the transformation, slow down further propagation of the crack, resulting in the enhancement of the mechanical properties [1,2]. Hence, this phase transformation is the key factor for obtaining materials with increased toughness and fatigue threshold.

On the other hand, the transformation can also be induced by environmental stresses, leading to the so-called ageing phenomenon [3,4]. The degradation resulting from this phenomenon is characterized by surface roughening, microcracking at the surface and particle release in the body [5]. Ageing of zirconia is indubitably the main factor limiting further development of Y-TZP as a biomaterial. In fact, several clinical failures of zirconia-based femoral heads have recently been reported. These 343 in vivo failures [6], associated with two particular batches of Prozyr® heads, were related to an unexpected acceleration of ageing at the surface due to a modification of processing conditions. Consequently, in absence of a complete and rational description and analysis of this ageing, there is a general trend of reverting back to the older materials solutions.

Ageing in zirconia was investigated by ceramists for the last two decades, and a number of influencing factors have been identified, the most important ones being grain size [7], density [7] and phase assemblage [8]. However, the recent failure events have clearly demonstrated there is room for additional factors to be assessed, since the involved materials satisfied the ISO requirements based on the existing known factors [9]. One of these factors, the role of residual stresses has been underestimated so far. However, as pointed out recently by Basu [10], 'It is evident that tensile residual stresses decrease the critical transformation stress and thereby increase the driving force for transformation by lowering the nucleation barrier, so contributing to enhance transformability of t-ZrO2'. Though some early studies suggested that, residual stresses were not relevant to ageing [11], other experimental reports clearly demonstrated the beneficial effects of relaxing residual stresses, for instance by the formation of a glassy phase at triple junctions resulting from oxide doping [12]. Since stresses theoretically play a key role in the transformation, residual stresses should have some effect on ageing. In fact, it was reported that the transformation always nucleates at triple junction in 3Y-TZP, a location where residual stresses are known to concentrate [13,14]. This effect of residual stress on the transformation propagation at the surface has been recently illustrated in ceria-doped zirconia, at the scale of the first transformed grains [15]. Furthermore, since ageing starts at the surface of the components, the type and magnitude of the residual stresses at surface should be of prime importance. The surface stress state is directly related to the machining and polishing conditions [16], which vary significantly with manufacturer/technique used. The ISO requirements for this aspect are very surprisingly inadequate, as it has never been fully investigated.

Therefore, the aim of the paper is to attempt a quantitative analysis of the role played by residual stresses and surface finish on the ageing sensitivity of biomedical grade 3Y-TZP at the macroscopic scale.

## 2. Experimental

### 2.1. Materials processing

Samples were processed from an atomised 3 mol% Y2O3 zirconia powder (TZ3Y, Tosoh, Tokyo, Japan), to obtain biomedical grade materials, according to ISO 13356:1997. Samples were sintered 5 h at 1450 °C, with heating and cooling rates of 300 °C/h. The microstructure of the material is illustrated in Fig. 1. The grain size (intercept segment length) and density were 0.4 μm and 6.08 g/cm3, respectively. Toughness and hardness values, measured via 10 kg indentations, were 6 MPa√m and 13 GPa, respectively.

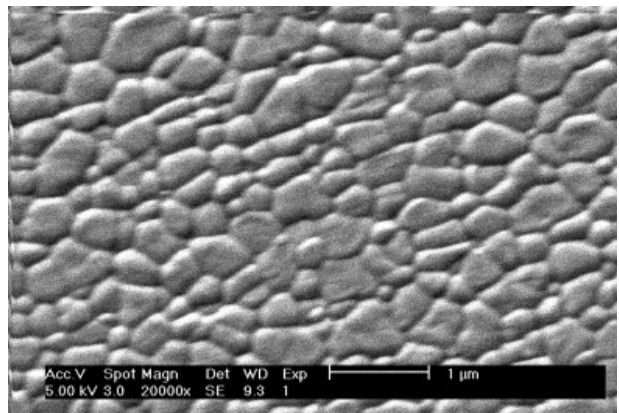

Fig. 1. Microstructure of the 3Y-TZP ceramic investigated in the present work (Scanning Electron Microscopy image).

### 2.2. Low-temperature autoclave ageing

The transformation is known to be both thermally activated and accelerated by the presence of water. Samples were hence placed in an autoclave (Fisher Bioblock Scientific, Illkirch, France) in steam for controlled time periods at 140 °C and under 3 bar pressure, in order to artificially induce the phase transformation at the surface. Since, the thermal activation (~106 kJ/mol) of the ageing process is known [17], it was possible to estimate that 20 min of such a treatment corresponded roughly to 1 year in vivo.

## 2.3. X-ray diffraction

The transformation, which is the underlying cause of the ageing phenomenon, was followed by measuring the phase fraction evolution using X-ray diffraction (XRD). XRD data were collected with a θ–2θ diffractometer (Rigaku, Tokyo, Japan) using Cu-$K_\alpha$ radiation. Diffractograms were obtained from 27° to 33°, at a scan speed of 0.2°/min and a step size of 0.02°. The monoclinic phase fraction, $X_m$, was calculated using the Garvie and Nicholson method [18]:

$$I_m = \frac{I_m(-111) + I_m(111)}{I_m(-111) + I_m(111) + I_t(101)} \quad (1)$$

where $I_t$ and $I_m$ represent the integrated intensity (area under the peaks) of the tetragonal (1 0 1) and monoclinic (1 1 1) and (−1 1 1) peaks. The monoclinic volume fraction, $V_m$, is then given by Toraya et al. [19]:

$$V_m = \frac{1.311\ X_m}{1 + 0.311\ X_m} \quad (2)$$

## 2.4. Optical microscopy

Optical microscopy in Normarsky contrast (Zeiss Axiophot, Germany) was used to investigate the surface degradation kinetics at the mesoscopic and microscopic scales. It is possible to observe the change in surface relief due to the t–m transformation, since it is accompanied by a lattice volume increase. It has been shown that the growth of the monoclinic phase at the surface was observable when their apparent diameter reaches as little as a few microns, without any specific surface preparation [17].

## 2.5. Atomic force microscopy (AFM)

AFM experiments were carried out with a D3100 nanoscope from Digital Instruments Inc., using oxide sharpened silicon nitride probes (Nanosensor, CONT-R model) in contact mode, with an average scanning speed of 10 µm/s. The vertical resolution of AFM allows for following the transformation-induced surface relief very precisely.

## 3. Results

## 3.1. Qualitative observations: preferential transformation around polishing scratches

It is plausible that some deep scratches introduced during machining and/or polishing operations, might influence the internal stress state even up to 20 µm below the surface [16], hence the ageing sensitivity. The influence of such scratches on the ageing degradation could be observed at the mesoscopic scale, as shown in Fig. 2. A partially transformed surface was observed by optical microscopy after ageing for 75 min (equivalent to 3.5 years in vivo). The formation of monoclinic spots at the surface is clearly observable. These spots subsequently grew in size (height and diameter) [17,20], in agreement with nucleation and growth models commonly used to describe phase transformations [21]. More interestingly, some line-shaped transformed zones up to a few millimetres long and only a few microns wide were also observed at the global scale. Further observations were performed at a more local scale using atomic force microscopy; a typical example is shown in Fig. 3. The AFM height image clearly reveals the various features of the surface relief, i.e. surface upheavals induced by the transformation (as observed in Fig. 2), as well as the scratches induced in the polishing stage, which appear as dark straight lines on the micrographs. From these experiments, it appears that monoclinic spots nucleate preferentially along the scratches. In addition, it is worth noting that the width of the transformed zone is almost constant along the scratches. Since the dimensions of the monoclinic spots vary linearly with time [17], it can be concluded that all these spots were formed at the same time along the scratches. Hence, it is quite clear that surface scratches act as favored nucleation sites. When the degradation continues, these spots grow and eventually coalesce altogether, leaving the observed elongated transformed zones, which reach lengths of up to a couple of millimetres, i.e. two orders of magnitude larger than the individual monoclinic spots appearing away from the scratches. It is also worth pointing out that some of the scratches observed in Figs. 2 and 3 did not lead to preferential transformation in their immediate surroundings. While the deepest scratches have the most deleterious effect on the transformation, the finest scratches were observed not to have a significant effect on the transformation.

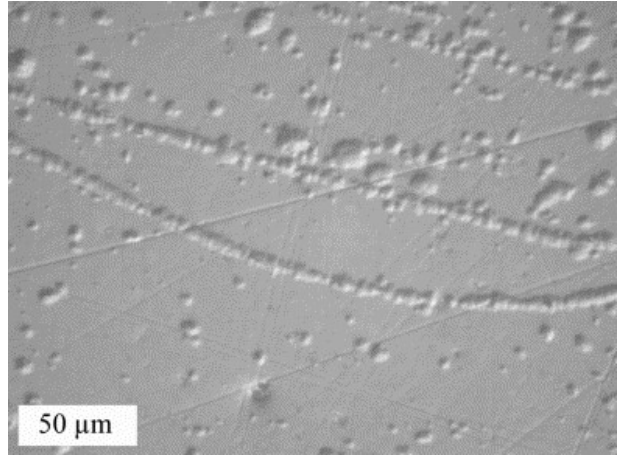

Fig. 2. Optical microscopy observation (Normarski contrast) of a partially transformed surface after 75 min at 140 °C, showing the initial stage of ageing.

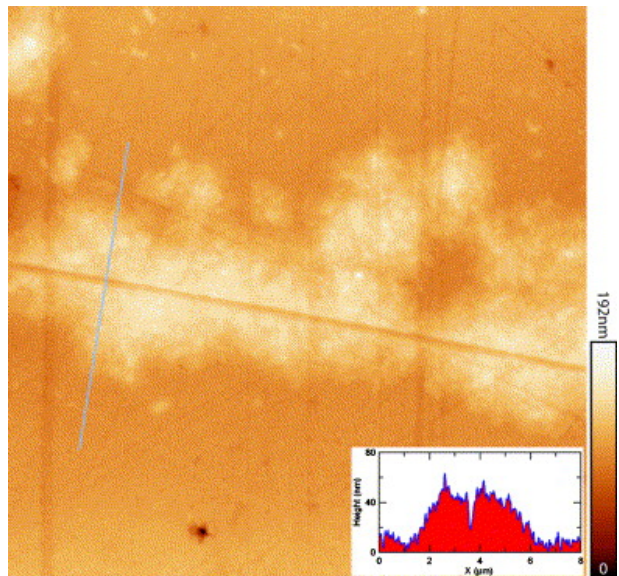

Fig. 3. Atomic force microscopy observation (height image) of a partially transformed surface, showing the preferential transformation (bright areas) in the surrounding of a polishing scratch. The inset represents a line profile perpendicular to the scratch.

The formation process of these elongated transformed zones was investigated by performing several consecutive autoclave treatments and observing the transformation propagation within the same area on the surface between each treatment steps. These observations, included in Fig. 4, plainly reveal the simultaneous, aligned, but independent formation of the spots in the initial stage. In the second stage, due to the growth of the transformed zones, they impinge on each other, leading to the formation of the homogeneously transformed lines previously identified. Since the transformation in autoclave is not induced by external mechanical stresses (as opposed to transformation around a propagating crack [13]), the grains are transformed as a function of their disequilibrium state, i.e. all the grains transforming at the same stage of the autoclave treatment are in a similar stress state. It can be inferred from these observations that when a scratch is created at the surface by machining or polishing, it leaves a homogeneous stress state in its surroundings. The areas affected by these stresses will later transform simultaneously during the ageing treatment. The presence of these scratches has, therefore, an extremely damaging influence on the ageing sensitivity as they act as nucleating agents and accelerate the surface degradation. This was further investigated to better assess such effects quantitatively, as detailed below.

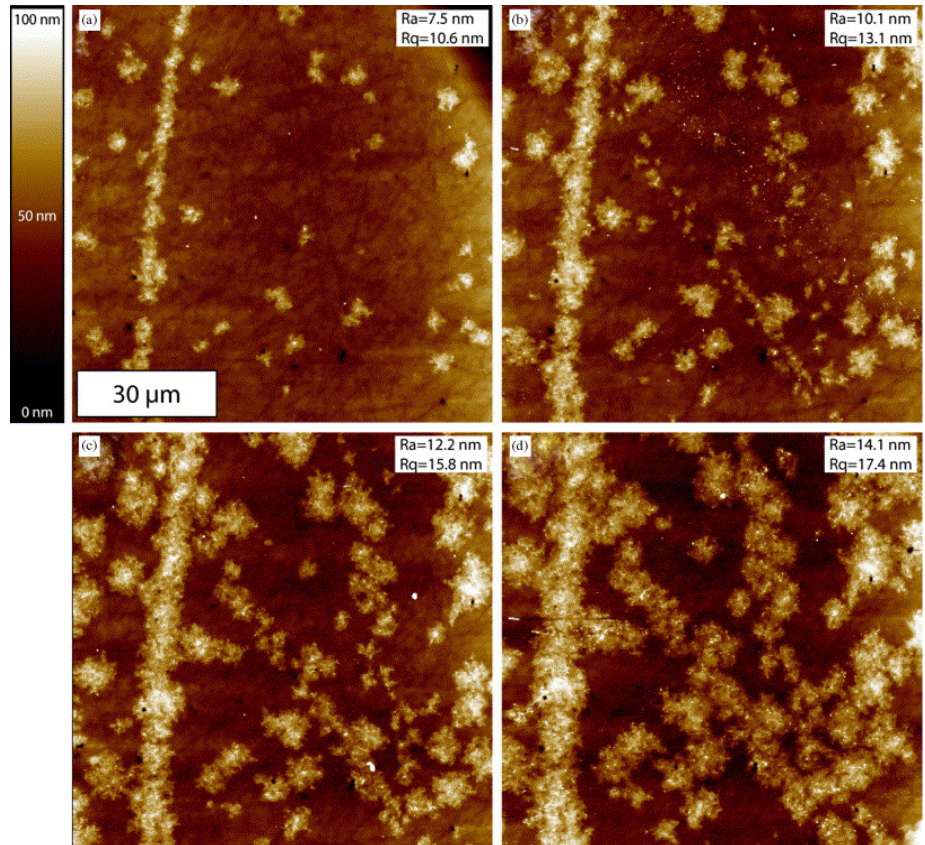

Fig. 4. AFM micrographs (height image) of the progressive nucleation and growth of monoclinic phase after (a) 40, (b) 60, (c) 80 and (d) 100 min at 140 °C.

## 3.2. Quantitative analysis: influence of roughness and residual internal stresses on ageing

An experimental procedure was designed to investigate quantitatively the influence of the surface preparation stages. A series of samples were processed under the same conditions, the only differences being the final polishing stages. All the samples were polished using standard diamond products, down to a 1 μm grade, to reach the same smooth surface state (roughness equal to 2 nm, as measured by AFM). Some samples were then "depolished" by using diamond pastes of larger grades (i.e. 3 and 6 μm), to intentionally generate surface scratches and consequently, modify the residual stresses induced by polishing. Half of the samples were subjected to a thermal treatment of 2 h at 1200 °C, known [22] to relax the residual stresses without affecting the surface relief. The samples preparation conditions and designation are given in Table 1, along with the ISO requirements in terms of surface state.

| Sample designation | Thermal treatment | Intermediate polishing grade (µm) | Final polishing grade (µm) | Roughness (Ra) (nm) |
|---|---|---|---|---|
| 1NT | No | 1 | 1 | 2 |
| 1T | 2 h at 1200 °C | 1 | 1 | 2 |
| 3NT | No | 1 | 3 | 3 |
| 3T | 2 h at 1200 °C | 1 | 3 | 3 |
| 6NT | No | 1 | 6 | 6 |
| 6T | 2 h at 1200 °C | 1 | 6 | 6 |
| ISO requirements | — | — | — | 30 |

Roughness was measured by AFM experiments. Roughness requirement from ISO 13356:1997.

Table 1. Sample designation and surface finish characteristics

The ageing kinetics of the samples are given in Fig. 5. The 1NT and 3NT samples exhibited exactly the same transformation kinetics; while the 6NT sample transformation rate was much lower. When subjected to the thermal treatment, all the samples exhibited the same transformation kinetics. Different residual stress states are reached as a function of the polishing stages, leading to different ageing sensitivity. Once the residual stresses are relaxed thermally, the transformation kinetics of all samples were practically identical, although the roughness was different.

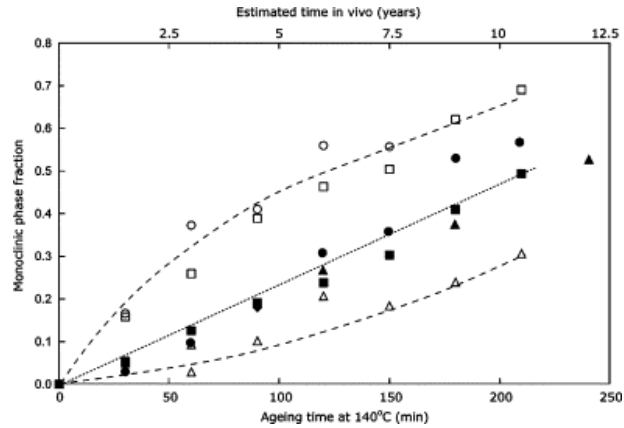

Fig. 5. Transformation kinetics measured at 140 °C by XRD for 1NT(○), 1T(●), 3NT (□), 3T (■), 6NT (∆) and 6T (▲) specimens (see Table 1 for sample designation).

4. Discussion

From these results, it can be first concluded that the residual stresses induced by scratches, more than their topography, are responsible for the influence of polishing stages on ageing. The stress state induced by the polishing stages can be further understood by analyzing the behavior of the different samples. For the 1 and 3 μm finish samples, it is obvious that the thermal treatment is favorable to the increase of stability—the transformation rate of the relaxed samples being lower than the as polished samples. Hence, the initial stress state must have a deleterious effect on the transformation sensitivity. Since a preliminary good finish was reached, it can be assumed that the 1 and 3 μm finishing stages only induced small defects at the surface, acting as preferential nucleation sites, which are relaxed after the thermal treatment. The samples polished with 6 μm finish exhibited a contrasting behavior, in that the thermal treatment had a deleterious effect on the ageing sensitivity. The transformation rate is increased when the thermal treatment is performed, which means the initial stress state (presumably, compressive stresses) was not favorable for the transformation to occur. The influence of machining and polishing on the residual stress state has been investigated in the past [16]. Rough polishing treatments were established to introduce a gradient of residual compressive stresses, the magnitude of the stresses being the largest at the surface.

Residual stresses arise from the elastic/plastic damage at the ceramic surface induced by the diamond grains. This is similar to the damage that occurs during the process of indentation, when a sharp diamond Vickers indenter causes a certain impression at the surface. A number of studies have looked to characterize and quantify the residual stress fields around Vickers indentations in ceramics. Fig. 6 shows a schematic cross section of a Vickers impression. When the indenter advances into the material, the impression is accommodated by plastic flow, resulting in an approximately hemispherical plastic zone around the indent surrounded by the confining elastic matrix. The plastic deformation process being volume-conservative, the indent volume is only accommodated by elastic strain. When the indenter is removed, the plastic zone inhibits elastic recovery from occurring completely, so that a residual stress field arises, associated with this elastic/plastic contact. The same kind of elastic/plastic contact and consequent residual stress field is likely to occur under a polishing diamond grain. The residual stresses are tensile in tangential directions outside the plastic zone, and hydrostatically compressive within the plastic region. The magnitude of the tensile residual stresses is a function of the applied load, the Young's modulus of the material and its hardness. It is maximum at the elastic/plastic interface and decreases with the distance to the indentation. These stresses were experimentally measured around 625 N indentation prints in a 3Y-TZP zirconia with the same microstructure and reached 200 MPa [22]. The magnitude of such tensile stresses can change the stability of zirconia grains, and the propensity to transform during ageing treatments. In the same work [22], some thermal treatments in air were performed after indentation, at different temperatures (from 200 to 1200 °C). At temperatures where the t–m transformation is fast, i.e. at 200 or 400 °C [3,4], a preferential transformation around the Vickers prints was noticed, in the area where the tensile stresses are large enough to trigger the transformation (cf. Fig. 6). This phenomenon accounts for the preferential transformation around polishing scratches in the case of 1 or 3 μm surface finish. This is illustrated schematically in Fig. 7a. In the case of the rough polishing (6 μm surface finish), compressive stresses must be present at the surface, since it is established that compressive stresses are less favorable for the transformation to occur than tensile stresses [23]. The density of scratches must be high enough to create a compressive stress field at the surface, just like if a series of indents was produced at the surface, close to each other. A schematic cross section of this configuration is given in Fig. 7b. This surface state is beneficial for the ageing resistance. These different stress states account for the stability of a given zirconia. These stresses can be relaxed, for instance by a thermal treatment at 1200 °C. Such a treatment was used here only to show the influence of residual stresses on ageing. It should not be applied to a ceramic joint head, since both compressive (in the cone) and tensile stresses (at the surface) should be present. More important is the care taken to minimize tensile residual stresses on the bearing surface, and to keep compressive stresses in the cone region.

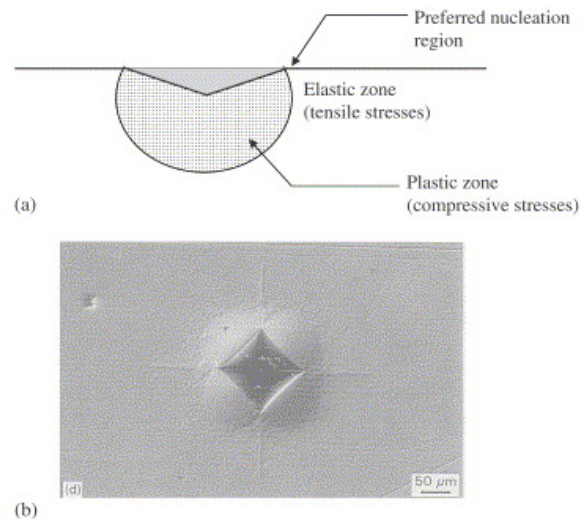

Fig. 6. (a) Schematic representation of elastic/plastic damage induced by an indentation and (b) preferential transformation around the indentation (from [22]).

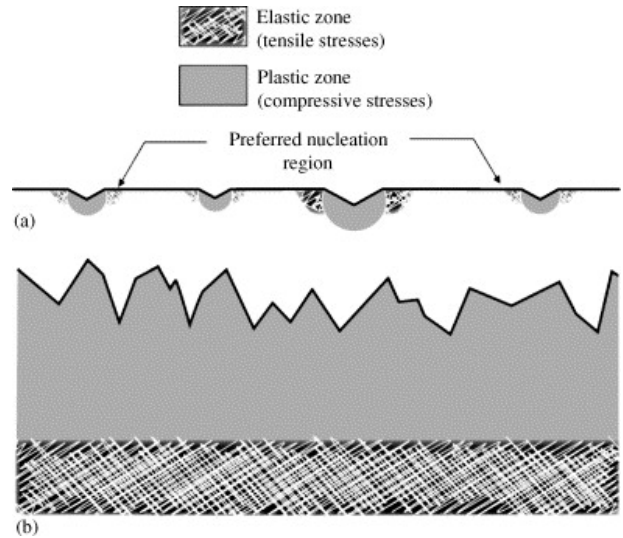

Fig. 7. (a) Schematic cross section representation of the material's surface for a 1 or 3 µm surface finish and (b) same representation for a larger density of surface scratches (6 µm surface finish).

If the surface long-term stability was the only parameter of interest, the presence of compressive residual stresses, associated with rough surfaces could be of interest. Surface states requirements are nonetheless of prime importance as far as bearing surfaces in total hip replacement are concerned. In order to minimize the wear phenomenon and particle release in the body, keeping a surface roughness as low as possible has always been an obvious objective for hip prostheses manufacturers. The requirement (Ra<30nm) introduced with the ISO13356 in 1997 was obviously inadequate to avoid excessive wear. Currently manufactured femoral heads and cups exhibit surface roughness value lying typically around 2–6 nm [24]. By keeping the surface roughness around these values, it was assumed this would be a criterion sufficient to ensure long-term stability. From the results presented here, it is quite clear that the entire machining and polishing process has a great influence on the ageing sensitivity. Variations of the surface state well below the ISO requirements produce materials with drastically different degradation kinetics. It appears that the actual ISO requirements concerning the long-term stability and surface state must be updated in light of the observations of the ageing phenomenon presented here.

## 5. Conclusions

The critical influence of polishing stages on the ageing sensitivity of 3Y-TZP has been systematically investigated by optical microscopy, atomic force microscopy and X-ray diffraction. The ageing sensitivity of biomedical grade zirconia is directly linked to the type (compressive or tensile) and amount of residual stresses. Rough polishing produces a compressive surface stress layer beneficial for the ageing resistance, while smooth polishing produces preferential transformation nucleation around scratches, due to elastic/plastic damage tensile residual stresses.

It appears finally that the actual ISO requirements concerning the long-term stability and surface state should be updated in regards of the actual knowledge of the ageing phenomenon, to ensure the long-term stability and success of Y-TZP biomedical components. The surface roughness cannot be used as the single indicator for ensuring in vivo stability.

## Acknowledgements

Financial support of the Rhône-Alpes region and the European Union (GROWTH2000, Project BIOKER, Reference GRD2-2000-25039) are acknowledged. The authors would also like to thank the Consortium de Laboratoires pour l'Analyse par Microscopie à Sonde Locale (CLAMS) for using the nanoscope.